\documentclass[12pt]{article}
\usepackage{graphicx}
\begin{document}

\pagestyle{empty}

\noindent
{\bf Do the Pulsation Properties of Red Giants Vary around their Long Secondary Period Cycle?}

\bigskip

\noindent
{\bf John R. Percy\\Department of Astronomy and Astrophysics, and\\Dunlap Institute of Astronomy and Astrophysics\\University of Toronto\\Toronto ON\\Canada M5S 3H4}

\medskip

\noindent
{\bf Kunming Di\\Department of Astronomy and Astrophysics\\University of Toronto\\Toronto ON\\Canada M5S 3H4}

\bigskip

{\bf Abstract}  We have used visual and Johnson V observations from the
American Association of Variable Star Observers (AAVSO) International Database,
and the AAVSO VSTAR time-series package, and (O-C) analysis to investigate possible changes in the
pulsation period and amplitude of the pulsating red giants U Del, EU Del, 
X Her, and Y Lyn around the cycle of their long secondary periods (LSPs).  We find
no such changes in period or amplitude.  This suggests (weakly) that the process which causes the
long secondary periods does not change the physical properties of the
visible hemisphere of the stars significantly as the LSP cycle proceeds.

\medskip

\noindent
AAVSO keywords = AAVSO International Database; Photometry, visual; pulsating variables; giants, red; period analysis; amplitude analysis

\medskip

\noindent
ADS keywords = stars; stars: late-type; techniques: photometric; methods: statistical; stars: variable; stars: oscillations

\medskip

\noindent
{\bf 1. Introduction}

\smallskip

About a third of pulsating red giants (PRGs) show a long secondary 
period (LSP),
5 to 10 times longer than the pulsation period, depending on the
pulsation mode (Wood 2000, Percy and Bakos 2003).  The cause
of the LSP is uncertain.  This is one of several unexplained phenomena
in PRGs.  The purpose of this short paper is to
investigate whether the pulsation
period and/or the amplitude changes around the LSP cycle.  

We do
this by examining four stars -- U Del, EU Del, X Her, and Y Lyn -- which
have well-determined LSPs, and which have precise photoelectric/CCD observations as well as visual
observations in the database of the American Association of Variable
Star Observers (AAVSO).  The pulsation properties of the four stars are
given in Table 1, which lists the pulsation period P in days, its
visual semi-amplitude, the LSP in days, and its visual semi-amplitude.
They are average
values; it is known that the pulsation periods and the LSPs and
their amplitudes ``wander"
with time.

\medskip

\noindent
{\bf 2. Data and Analysis}

\smallskip

We used visual and Johnson V (PEP or CCD) observations from the AAVSO International Database (AID:
Kafka 2018), the AAVSO VSTAR time-series analysis package (Benn 2013)
which includes both a Fourier analysis and a wavelet analysis routine,
and (O-C) analysis to study the periods and amplitudes of the four PRGs.

\medskip

\clearpage

\noindent
{\bf 3 Results}

\medskip

\noindent
{\bf 3.1 Previously-Known Time Scales of Period and Amplitude Variation}

\smallskip

It is already known, from the study of dozens of stars, that the periods of PRGs vary or ``wander" on time scales
of about 40 pulsation periods (Eddington and Plakidis 1929,
Percy and Colivas 1999, Percy and Qiu 2018), and the amplitudes of PRGs vary on time
scales of 20-35 pulsation periods (Percy and Abachi 2013, Percy and Laing 2017).  This is
strong evidence that the pulsation periods and amplitudes do {\it not} vary primarily
on the same time scale as the LSPs, which are only 5-10 pulsation
periods in length.

\medskip

\noindent
{\bf 3.2 Specifically-Determined Time Scales of Period and Amplitude Variations}

\smallskip

We used the wavelet routine in VSTAR to study the change in period and
amplitude in the last three LSP cycles in the four program stars, to see
whether there might be three cycles of period and/or amplitude variation.  The
results are given in Table 2, which lists the star, the JD range (minus
2400000), and a description of the behavior of the pulsation period and amplitude.

In no case does the period or amplitude vary on the LSP time scale.
The variations are much slower than the LSP, and consistent
with the time scales given in section 3.1.

\smallskip

\begin{figure}
\begin{center}
\includegraphics[height=7cm]{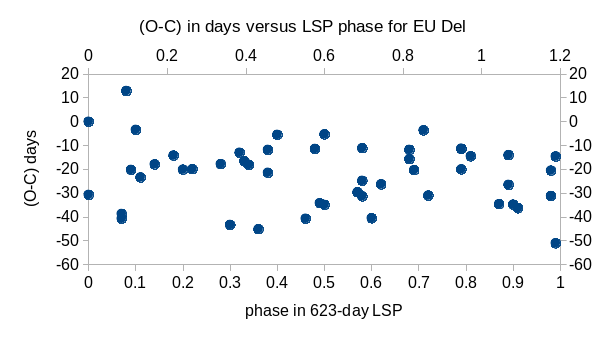}
\caption{The pulsational (O-C) values for EU Del, versus the phase in the long
secondary period.  The LSP is 624.2 days, and t(0) is JD 2447450.
There is no evidence that the (O-C) values, and hence the pulsation period
depend on the LSP phase.}
\end{center}
\end{figure}

\begin{figure}
\begin{center}
\includegraphics[height=7cm]{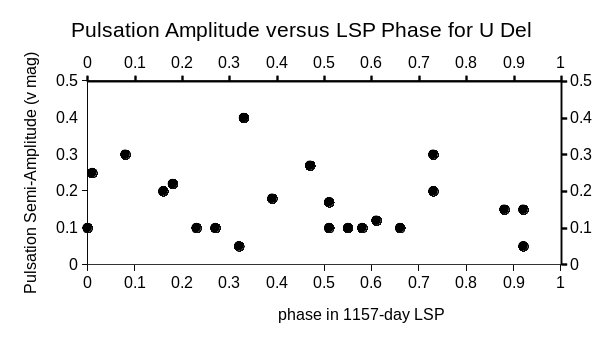}
\caption{The pulsation semi-amplitude for U Del, determined from PEP/CCD
data, versus the phase  in the long secondary period.  There is no
evidence that the pulsation amplitude depends on the LSP phase.  The same
is true for EU Del and X Her (Table 2).}
\end{center}
\end{figure}


\medskip

\noindent
{\bf 3.3 (O-C) Analysis of Period Variations versus LSP Phase}

\smallskip

(O-C) analysis (Percy 2007, p. 68) is an alternative method for studying
period changes with time.  Our stars have precise Johnson V observations which
can be used to determine times of maximum in the pulsation cycle.  We also
have accurate values of the pulsation period and LSP.

EU Del is best suited for this analysis, and will be used to demonstrate
it.  Its pulsation period and LSP are
especially well-determined.  It has extensive photometry which can be used to determine
times of pulsation maximum.  Table 3 lists times of pulsation maximum, minus 2400000, their
(O-C) values, and their LSP phase, using the first pulsation maximum
as t(0).

Figure 1 shows a plot of pulsation (O-C) versus LSP phase.  There is no correlation;
Figure 1 is a scatter diagram.  Similar results were obtained for the other
three stars.  There was no correlation between (O-C) and LSP phase.  (O-C)s
during a single LSP cycle were either constant within the observational
error, or varied slowly
as part of the longer timescale variability discussed above.  

The
analysis was complicated, in some cases, by the fact that the pulsation amplitude sometimes decreased
to close to zero at some epochs as part of its long-term variation.  It was therefore difficult to
identify times of pulsation maximum around those times.

\medskip

\noindent
{\bf 3.4 Pulsation Amplitude versus LSP Phase}

\smallskip

U Del, EU Del, and X Her have series of Johnson V observations which can be
used to estimate the pulsation amplitude.  Table 4 lists the {\it semi-amplitudes}
(to be consistent with the output of VSTAR) and the mean JDs (minus 2400000) of such
series.  In none of the three stars is there a relation between pulsation
amplitude and LSP phase.  Figure 2 shows one example: for U Del.  The
graph of pulsation semi-amplitude {\it versus} LSP phase is a scatter diagram.  On the other hand: Table 4 shows that the
pulsation amplitude varies more slowly, on a time scale of several LSPs (20-35 pulsation periods), consistent with the discussion 
in section 3.1.

\medskip

\noindent
{\bf 4. Discussion}

\smallskip

There is no generally-accepted and proven explanation for the ``wandering"
pulsation periods, variable amplitudes, and LSPs in
pulsating red giants.  The wandering pulsation periods have been modelled as
random cycle-to-cycle fluctuations.  These might be connected with random
convective motions.  Large convective cells have long been predicted
to occur in the outer layers of PRGs, and have recently been imaged
on a PRG (Paladini {\it et al.} 2018).  The turnover times and rotational-variability times of these are expected
to be much longer than the pulsation period; see Percy and Deibert (2016) for
a discussion.

One might expect LSP-phase-dependent variations in pulsation period and
amplitude if the LSP was produced by significant changes in the physical properties of the
stellar surface facing the observer.  This would include the turnover of
large convective cells, or the rotation of large convective cells onto and off the
visible hemisphere.  On the other hand: if the LSP was due to a well-detached binary
companion, or to a large cloud of dust, orbiting the star, or any other
external process, then the stellar
surface and its pulsation properties would not be expected to change significantly.  The
binary scenario is discussed by Soszy\'{n}ski and Udalski (2014).  EG And is
a specific example.  It has a pulsation period of 29 days and an LSP of 242 days 
(Percy {\it et al.} 2001).  The LSP is exactly half of the orbital period of 483.3 days (Kenyon
and Garcia 2016).  Many other PRGs are likely to have binary companions.


\medskip

\noindent
{\bf 5. Conclusions}

\smallskip

From a study of four well-studied pulsating red giants with
LSPs, we find no evidence that the pulsation period and amplitude vary
with LSP phase.  Both vary on much longer (and different) time scales.
Our results suggest, weakly, that the LSP process -- whatever it is -- does not produce significant
variations in the physical properties of the visible hemisphere of the star.

\medskip

\noindent
{\bf Acknowledgements}

\smallskip

We thank the AAVSO observers who made the observations on which this project
is based, the AAVSO staff who archived them and made them publicly available, and the developers
of the VSTAR package which we used for analysis.
This project made
use of the SIMBAD database, maintained in Strasbourg, France.
The Dunlap Institute is funded through an endowment established by the David Dunlap family and the University of Toronto.

\bigskip

\noindent
{\bf References}

\smallskip

\noindent
Benn, D. 2013, VSTAR data analysis software {http://www.aavso.org/vstar-overview)

\noindent
Eddington, A.S., and Plakidis, S. 1929, {\it Mon. Not. Roy. Astron. Soc.}, {\bf 90}, 65.

\noindent
Kafka, S. 2018, variable star observations from the AAVSO International Database

(https://www.aavso.org/aavso-international-database)

\noindent
Kenyon, S.J., and Garcia, M.R. 2016, {\it Astrophys. J.}, {\bf 152}, 1.

\noindent
Paladini, C. {\it et al.}, 2018, {\it Nature}, {\bf 553}, 310.

\noindent
Percy, J.R., and Colivas, T. 1999, {\it Publ. Astron. Soc. Pacific}, {\bf 111}, 94.

\noindent
Percy, J.R., Wilson, J.B., and Henry, G.W. 2001, {\it Publ. Astron. Soc. Pacific}, {\bf 113}, 983.

\noindent
Percy, J.R., and Bakos, G.A., 2003, in {\it The Garrison Festschrift}, ed.
R.O. Gray, C. Corbally, and A.G.D. Philip, L. Davis Press, Schenectady NY, 49.

\noindent
Percy, J.R., and Abachi, R. 2013, {\it J. Amer. Assoc. Var. Star Obs.}, {\bf 41}, 193.

\noindent
Percy, J.R., and Deibert, E. 2016, {\it J. Amer. Assoc. Var. Star Obs.}, {\bf 44}, 94. 

\noindent
Percy, J.R., and Laing, J. 2017, {\it J. Amer. Assoc. Var. Star Obs.}, {\bf 45}, 197.

\noindent
Percy, J.R., and Qiu, A.L. 2018, arxiv.org/abs/1805.11027

\noindent
Soszy\'{n}ski, I., and Udalski, A. 2014, {\it Astrophys. J.}, {\bf 788}, 13.

\noindent
Wood, P.R. 2000, {\it Publ. Astron. Soc. Australia}, {\bf 17}, 18.

\medskip

\smallskip
\begin{table}
\begin{center}
\caption{Four Pulsating Red Giants with Long Secondary Periods}
\begin{tabular}{rrccc}
\hline
Star & P days & A mag & LSP days & A(LSP) mag \\
\hline
U Del & 117.8 & 0.05 & 1157.4 & 0.20 \\
EU Del & 62.688 & 0.06 & 624.2 & 0.06 \\
X Her & 101.37 & 0.06 & 738.0 & 0.08 \\
Y Lyn & 134: & 0.08 & 1251.6 & 0.34 \\
\hline
\end{tabular}
\end{center}
\end{table}

\medskip

\begin{table}
\begin{center}
\caption{Pulsation Period and Amplitude Variations in Pulsating Red Giants}
\begin{tabular}{rrll}
\hline
Star & JD range & P variation & A variation \\
\hline
U Del & 54788-58260 & discontinuous & monotonic decrease  \\
EU Del & 56388-58260 & discontinuous & one cycle \\
X Her & 56046-58260 & half-cycle & monotonic decrease \\
Y Lyn & 54504-58260 & discontinuous & half-cycle \\
\hline
\end{tabular}
\end{center}
\end{table}

\begin{table}
\begin{center}
\caption{Pulsation Semi-Amplitudes (V) versus LSP Phase for Three Stars}
\begin{tabular}{rrr|rrr|rrr}
\hline
\multicolumn{3}{c}{U Del} & \multicolumn{3}{c}{EU Del} & \multicolumn{3}{c}{X Her} \\
\hline
JD & A & N.phase & JD & A & N.phase & JD & A & N.phase \\
\hline
48180 & 0.10 & 0.00 & 47400 & 0.35 & 0.30 & 56475 & 0.43 & 0.95 \\
48550 & 0.05 & 0.32 & 47780 & 0.25 & 0.91 & 56875 & 0.21 & 1.49 \\
48885 & 0.12 & 0.61 & 48175 & 0.25 & 1.54 & 57175 & 0.37 & 1.89 \\
49250 & 0.15 & 0.92 & 48525 & 0.30 & 2.10 & 57250 & 0.33 & 2.00 \\
49650 & 0.10 & 1.27 & 48920 & 0.35 & 2.73 & 57310 & 0.35 & 2.08 \\
49975 & 0.10 & 1.55 & 49250 & 0.25 & 3.26 & 57575 & 0.30 & 2.44 \\
50400 & 0.05 & 1.92 & 49650 & 0.20 & 3.90 & 57900 & 0.20 & 2.88 \\
51080 & 0.10 & 2.51 & 50000 & 0.21 & 4.46 & 57950 & 0.25 & 2.94 \\
51840 & 0.20 & 3.16 & 50350 & 0.23 & 5.02 & 58225 & 0.15 & 3.32 \\
52200 & 0.27 & 3.47 & 50700 & 0.15 & 5.58 & & & \\
52500 & 0.20 & 3.73 & 51100 & 0.18 & 6.23 & & & \\
52900 & 0.30 & 4.08 & 51450 & 0.25 & 6.79 & & & \\
53260 & 0.18 & 4.39 & 51825 & 0.18 & 7.39 & & & \\
53650 & 0.30 & 4.73 & 52175 & 0.15 & 7.95 & & & \\
53975 & 0.25 & 5.01 & 52575 & 0.23 & 8.59 & & & \\
54350 & 0.40 & 5.33 & 52920 & 0.10 & 9.14 & & & \\
54725 & 0.10 & 5.66 & 53260 & 0.25 & 9.69 & & & \\
56550 & 0.10 & 7.23 & 54000 & 0.13 & 10.87 & & & \\
56950 & 0.10 & 7.58 & 54360 & 0.50 & 11.45 & & & \\
57300 & 0.15 & 7.88 & 55100 & 0.20 & 12.64 & & & \\
57650 & 0.22 & 8.18 & 56525 & 0.20 & 14.92 & & & \\
58025 & 0.17 & 8.51 & 56950 & 0.18 & 15.60 & & & \\
      &      &      & 57250 & 0.23 & 16.08 & & & \\
      &      &      & 58050 & 0.13 & 17.36 & & & \\
\hline
\end{tabular}
\end{center}
\end{table}

\begin{table}
\begin{center}
\caption{(O-C) versus LSP Phase for EU Delphini}
\begin{tabular}{ccr|ccr|ccr}
\hline
JD & (O-C) d & N.phase & JD & (O-C) d & N.phase & JD & (O-C) d & N.phase \\
\hline
47450 & 0.0 & 0.00 & 49560 & -21.4 & 3.38 & 52952 & -14.5 & 8.81 \\
47502 & +12.9 & 0.08 & 49622 & -22.1 & 3.48 & 53267 & -13.0 & 9.32 \\
47689 & -11.8 & 0.38 & 49682 & -24.8 & 3.58 & 53998 & -34.2 & 10.49 \\
47752 & -11.4 & 0.48 & 49937 & -20.5 & 3.98 & 54385 & -23.4 & 11.11 \\
47815 & -11.1 & 0.58 & 50000 & -20.2 & 4.09 & 54315 & -30.7 & 11.00 \\
47877 & -11.8 & 0.68 & 50304 & -29.6 & 4.57 & 55078 & -19.9 & 12.22 \\
48188 & -14.2 & 1.18 & 50376 & -20.3 & 4.69 & 55144 & -16.6 & 12.33 \\
48385 & -5.3 & 1.50 & 50439 & -20.0 & 4.79 & 55502 & -34.8 & 12.90 \\
48500 & -15.7 & 1.68 & 51114 & -34.6 & 5.87 & 55878 & -34.9 & 13.50 \\
48567 & -11.4 & 1.79 & 51417 & -45.1 & 6.36 & 56562 & -40.4 & 14.60 \\
48763 & -3.4 & 2.10 & 51484 & -40.7 & 6.46 & 56898 & -17.9 & 15.14 \\
48874 & -17.8 & 2.28 & 51749 & -26.5 & 6.89 & 56998 & -43.3 & 15.30 \\
48949 & -5.5 & 2.40 & 51807 & -31.2 & 6.98 & 57203 & -26.3 & 15.62 \\
49139 & -3.6 & 2.71 & 51860 & -40.7 & 7.07 & 57261 & -31.0 & 15.72 \\
49194 & -11.3 & 2.79 & 52183 & -31.3 & 7.58 & 57650 & -18.1 & 16.34 \\
49254 & -14.0 & 2.89 & 52489 & -38.7 & 8.07 & 58008 & -36.3 & 16.91 \\
49316 & -14.6 & 2.99 & 52570 & -20.4 & 8.20 & 58056 & -51.0 & 16.99 \\
\hline
\end{tabular}
\end{center}`
\end{table}

\end{document}